# Quantum Fourier Transform–Based Denoising: Unitary Filtering for Enhanced Speech Clarity

[1]Rajeshwar Tripathi, [2]Sahil Tomar, [3]Sandeep Kumar, [4]Monika Aggarwal
[1,2,3]CRL Bharat Electronics Limited Ghaziabad, [4]Indian Institute of Technology Delhi
[1]tripathirajeshwar402@gmail.com, [2]sahiltomar893@gmail.com, [3]sann.kaushik@gmail.com, [4]maggarwal@care.iitd.ernet.in

*Abstract—* This paper introduces a quantum-inspired denoising framework that integrates the Quantum Fourier Transform (QFT) into classical audio enhancement pipelines. Unlike conventional Fast Fourier Transform (FFT)-based methods, QFT provides a unitary transformation with global phase coherence and energy preservation, enabling improved discrimination between speech and noise. The proposed approach replaces FFT in Wiener and spectral subtraction filters with a QFT operator, ensuring consistent hyperparameter settings for fair comparison. Experiments on clean speech, synthetic tones, and noisy mixtures across diverse signal-to-noise ratio (SNR) conditions (−10 dB to +10 dB) demonstrate statistically significant gains in SNR, with up to 15 dB improvement and reduced artifact generation. Results confirm that QFT-based denoising offers robustness under low-SNR and nonstationary noise scenarios without additional computational overhead, highlighting its potential as a scalable pathway toward quantum-enhanced speech processing.

*Keywords— Quantum Fourier Transform, Audio Denoising, Wiener Filter, Spectral Subtraction, Signal-to-Noise Ratio.*

## I. INTRODUCTION

Denoising, the process of removing unwanted perturbations from signals, is a cornerstone challenge across a multitude of domains. In audio communication, background noise can render speech unintelligible; in biomedical monitoring, electrical and ambient interference can mask critical physiological patterns; in radar and imaging, clutter degrades detection accuracy; and in industrial sensing, mechanical vibrations and electromagnetic interference obscure sensor readings [1]. The ubiquity of noisy environments makes denoising an essential operation in day-to-day life as well as in high-stakes applications, underpinning everything from mobile telephony and hearing aids to medical diagnostics and autonomous navigation. Denoising aims to recover an estimate of the underlying clean signal from an observed mixture, where noise represents a broad spectrum of disturbances, including additive white Gaussian noise (AWGN), colored noise, impulse noise, and environmental interference etc. manifests in diverse forms, each with distinct statistical and spectral characteristics like AWGN serves as a foundational model for thermal and electronic noise, characterized by its constant power spectral density (PSD) across all frequencies [2]. In contrast, colored noise—encompassing pink (1/f) and brown (1/f²) noise— exhibits frequency-dependent power distributions and is commonly observed in natural phenomena and electronic systems [3][4]. Beyond these, impulse noise manifests as sparse, high-amplitude spikes caused by abrupt disturbances such as switching events or transmission errors [5], diverging sharply from the continuous nature of AWGN and colored noise. Additionally, environmental or broadband noise, such as café chatter, engine rumble, and wind, represent complex etc. represent non-stationary mixtures that challenge signal processing in real-world applications. Together, these noise types span a spectrum of statistical and spectral behaviors, reflecting diverse sources and contexts in both engineered and natural systems. Each noise type poses distinct challenges for separation, requiring adaptive strategies that consider both temporal and spectral characteristics. Widely adopted classical techniques include Wiener filtering and spectral subtraction, which act in the frequency domain to estimate clean signals from noisy observations [6]. These methods minimize error or subtract noise PSD to enhance audio clarity. However, these methods face significant challenges: Spectral subtraction often introduces musical noise artifacts, which manifest as perceptually intrusive tonal distortions, particularly in low signal-to-noise ratio (SNR) environments [7]. Meanwhile, Wiener filters depend heavily on precise estimation of noise and signal PSDs, yet their performance degrades sharply under rapidly changing or non-stationary noise conditions [6]. Compounding these issues, many approaches suffer from non-adaptive tuning, where fixed thresholds struggle to balance competing demands in mixed-noise scenarios. For instance, spectral subtraction relies on static over-subtraction factors (e.g., $\alpha$) to suppress noise. While effective in stationary noise environments, these fixed parameters fail to adapt to dynamic conditions (e.g., sudden changes in noise type or SNR), often leading to over-suppression of speech components (e.g., high-frequency consonants) or inadequate noise reduction in transient noise bursts [6]. Similarly, Wiener filtering's gain function $G(\omega)$ depends on fixed signal and noise PSD estimates. When noise characteristics shift rapidly (e.g., in a café with fluctuating chatter), static PSD estimates cause the filter to either blur speech details (by over-attenuating weak components) or retain residual noise (by under-attenuating dominant noise bins) [6]. This rigidity manifests in two critical failure modes i.e. Excessive Speech Distortion and Inadequate Noise Suppression. Together, these limitations highlight the need for adaptive, context-aware denoising strategies that mitigate artifacts while maintaining robustness across dynamic noise conditions. Recent studies highlight these limitations: implicit Wiener methods marginally outperform spectral subtraction but still struggle in dynamic acoustic environments [8]; jointly using Wiener filters and Deep Neural networks (DNN) improves results yet remains highly data-dependent and complex [9]. Adaptive thresholding in wavelet domains can suppress musical noise but relies on parameter tuning and may distort fine signal details [10]. Modern denoising heavily leans on deep learning (DL) models, including DNNs, subband masking, and denoising autoencoders. These yield impressive gains in intelligibility and perceptual quality [11], but demand large, annotated datasets, suffer from domain-generalization issues, and often operate as opaque black boxes—limits for interpretable, low-latency, or resource-constrained applications.

Quantum computing has surged forward as a "technology of the future," promising fundamentally new algorithmic paradigms. In signal processing, initial quantum inspired methods (e.g., using tensor networks or quantum inspired optimization) have shown potential for high dimensional data analysis [12]. While quantum-enhanced filtering has garnered exploratory interest, existing approaches face critical limitations that hinder practical applicability. Early work leveraging quantum principal component analysis demonstrated potential for isolating signal subspaces [13], yet these frameworks lack integration with realistic noise models, limiting their utility in real-world scenarios. Similarly, quantum-inspired annealing techniques have been applied to optimize adaptive filter coefficients [14], but their reliance on heuristic parameter embedding and poor scalability with increasing problem complexity restrict their effectiveness. Other studies propose quantum-inspired denoising strategies, such as amplitude-thresholding and spectral filtering via Walsh-Hadamard transforms [15] [16], yet these methods remain largely theoretical, with minimal empirical validation on real-world audio signals. Collectively, these efforts highlight the nascent stage of quantum-enhanced filtering, underscoring the need for robust, scalable frameworks that bridge theoretical constructs with practical noise environments. These reviews also had their limitations of reliance on classical Fourier bases, Lack of comprehensive evaluation etc. From all these reviews it is quite clear that many limitations remain in classical methods and Deep Learning methods. Despite theoretical promise, applying Quantum Fourier transform (QFT) based bases in practical denoising filters is understudied. Past works focus on transform implementation efficiency, not on denoising performance or real-data evaluation. well as quantum methods. So, quantum methods are left to be explored widely as well.

This study addresses these gaps by proposing the injection of a QFT unitary into classical denoising pipelines, hypothesizing that the richer phase–amplitude separation afforded by QFT can enhance noise–signal discrimination under challenging noise conditions. The novel contributions in this study are:

- This work pioneers the application of QFT to classical audio denoising pipelines, replacing heuristic Fast Fourier Transform (FFT) based spectral analysis with a unitary quantum-inspired framework. Unlike prior quantum signal processing studies limited to theoretical or synthetic domains, this study demonstrates empirical feasibility and performance gains in real-world noise suppression tasks.
- By leveraging QFT's inherent energy preservation and global phase coherence, proposed method avoids the spectral leakage and discontinuities inherent in classical windowed transformations. This results in superior noise-floor control and reduced artifact generation (e.g., musical noise), particularly in low-SNR regimes where classical spectral subtraction falters.
- A systematic empirical validation framework is introduced to compare classical and quantum Fourier transform (QFT)-based denoising pipelines. Identical hyperparameter configurations ($\alpha = 1.5$, $\beta = 0.02$) are enforced across both methodologies, applied post-transformation to ensure direct comparability, confounding variables are eliminated, and the transformation basis isolated as the critical differentiator. This rigorous approach reveals that QFT's unitary structure and global energy redistribution yield statistically significant improvements in SNR gain consistency and algorithmic stability, particularly under low-SNR conditions where classical spectral subtraction exhibits heightened sensitivity to heuristic parameter choices

The remainder of this paper is organized as follows: The Background section reviews classical denoising filters, including Wiener filtering and spectral subtraction, and introduces the theoretical framework of the QFT. Following this, the Methodology section details the integration of QFT into classical filtering pipelines, outlining the quantum-inspired denoising framework. The Experiment section outlines data preparation protocols, implementation details, and evaluation metrics used to assess performance. Subsequently, the Results and Discussion section presents empirical findings, comparing the QFT based approach with classical methods in terms of SNR improvement and artifact reduction. Finally, the Conclusion section summarizes key insights, highlights the contributions of this work, and proposes directions for future research in quantum-enhanced signal processing.

## II. BACKGROUND

### A. Denoising Pipeline

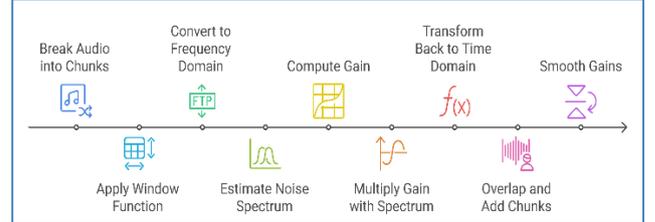

Fig. 1. Denoising workflow of audio

A detailed, step-by-step description of the classical denoising pipeline, highlighting the key equations at each stage, is presented in Fig. 1. Each step is detailed as follows: We first chop the continuous time-domain signal $x[n]$ into overlapping frames ("chunks") of length $L$, with a fixed hop (stride) of $H$ samples into $x_m[n]$:

$$x_m[n] = x[mH + n], n = 0, \ldots, L-1, m = 0,1,2,\ldots \quad (1)$$

The overlap between consecutive frames is determined by the hop size $H$. A typical choice is $H = L/2$ (50% overlap), which ensures smooth transitions between frames and avoids discontinuities in the reconstructed signal [17][18]. Overlapping compensates for the attenuation of signal energy at frame edges caused by windowing. Without overlapping, abrupt transitions between frames would reintroduce artifacts like musical noise [17][18]. To mitigate spectral leakage at the frame boundaries, each chunk computed from eq. (1). is multiplied by a smooth window $w[n]$ (e.g. Hamming, Hann):

$$x_m[n] \leftarrow x_m[n]w[n], w[n] \in [0,1] \quad (2)$$

This windowing operation tapers the edges of each frame, reducing abrupt amplitude changes that could distort the frequency-domain representation. Short-time Fourier transform (STFT) of each windowed frame from eq. (2). is computed to get its complex spectrum shown below:

$$Y_m(k) = \sum_{n=0}^{L-1} x_m[n]\, e^{-j2\pi kn/L}, k = 0, \ldots, L-1. \quad (3)$$

Here, the STFT decomposes the signal into time-frequency bins, enabling noise suppression in the spectral domain. Either the average magnitude ($\widehat{N}(k)$) or PSD ($\widehat{\phi}_N(k)$) of the noise is accumulated during non-speech (or via a minimum-statistics tracker) which is shown below:

$$\widehat{N}(k) = \frac{1}{\mu}\sum_{m \in \mu} |Y_m(k)|,$$

or,

$$\widehat{\phi}_N(k) = \frac{1}{\mu}\sum_{m \in \mu} |Y_m(k)|^2,$$

(4)

A real-valued "gain" $G_m(k) \in [0,1]$ is calculated for each time–frequency bin to suppress noise based on filter. This gain is derived from the estimated noise statistics $\widehat{N}(k)$ or $\widehat{\phi}_N(k)$ in eq. (4) and the noisy spectrum $Y_m(k)$ in eq. (3). The gain function acts as a frequency- and time-dependent attenuation factor, reducing the magnitude of noisy components while preserving the phase of the original signal. For instance, in Weiner filter

$G_m(k) = \frac{|\hat{s}_m(k)|^2}{|\hat{s}_m(k)|^2 + |\hat{N}_m(k)|^2}$, where $\hat{s}_m$ and $\hat{N}_m(k)$ are estimates of the clean signal and noise spectra, respectively [1].And then the gain is applied to the noisy spectrum (keeping the original phase) depicted below :

$$\hat{S}_m(k) = G_m(k) Y_m(k) \quad (5)$$

Here $Y_m(k)$ represents the Noisy complex spectrum (input to the filter), $\hat{S}_m(k)$ represents Denoised complex spectrum (output of the filter). The phase of $Y_m(k)$ is retained to avoid introducing phase distortion, which can degrade perceptual quality [19]. Each modified spectrum is then inverted via the inverse DFT.

$$\hat{s}_m[n] = \frac{1}{L}\sum_{k=0}^{L-1} \hat{S}_m(k) e^{j2\pi kn/L} \quad (6)$$

The full denoised signal is then reconstructed as in eq. (7) by overlapping-adding the time-domain frames using the same hop $H$ and window $w[n]$:

$$\hat{s}[n] = \sum_m \hat{s}_m[n - mH] w[n - mH] \quad (7)$$

This overlap-add method ensures seamless recombination of frames. The choice of window function (e.g., Hann) and hop size $H$ is critical. The overlap-add method reconstructs the signal by summing the tapered frames at their overlapping regions. For perfect reconstruction, the window $w[n]$ must satisfy the Constant Overlap-Add (COLA) condition as follows:

$$\sum_m w[n - mH] = 1, \forall n \quad (8)$$

Common windows like Hann ($w[n] = 0.5 - 0.5\cos(2\pi n/L)$) satisfy COLA for $H = L/2$, ensuring no amplitude distortion in the final signal [18]. Residual "musical noise" is reduced often by applying either temporal or spectral smoothing to the gain surface. Together, these steps form the backbone of virtually all classical spectral domain denoising algorithms.

## B. Classical Filter

*1) Weiner Filter:* A Wiener filter is an optimal linear time-invariant filter that produces an estimate $\hat{s}(n)$ of a desired random process $s(n)$ from a noisy observation $y(n) = s(n) + n(n)$ by minimizing the mean-square error (MSE) between the estimate and the true signal represented below:

$$G(\omega) = \frac{S_{sy}(\omega)}{S_{yy}(\omega)} = \frac{\Phi_S(\omega)}{\Phi_S(\omega) + \Phi_N(\omega)} \quad (9)$$

where $\Phi_S$, $\Phi_N$ are the power spectral densities of signal and noise. The Wiener filters each frequency bin using the a-priori SNR equated as:

$$G(\omega) = \frac{SNR(\omega)}{1 + SNR(\omega)} \quad (10)$$

effectively suppressing noise-dominated frequencies and retaining those dominated by the signal. The Wiener filter offers a Minimum mean square error (MMSE) optimal, linear estimation framework that minimizes mean-square error by weighing each frequency bin according to its estimated SNR. This results in a balanced reduction in noise and retention of speech. It adapts naturally to spectral variations and avoids the tonal artifacts associated with subtraction-based methods. On the downside, it requires precise estimates of signal and noise PSDs, which can be impractical in real-world settings. However, the Wiener filter has notable limitations. Its smooth gain function, while effective at reducing abrupt transitions that cause musical noise, tends to over-suppress weaker speech components. This occurs because the gain function attenuates frequencies proportionally to the estimated noise power $\Phi_S(\omega)$, even when low-energy speech components (e.g., unvoiced consonants or transient sounds) are present. For instance, in bins where $\Phi_S(\omega) \approx \Phi_N(\omega)$, the gain $G(\omega) \approx 0.5$, reducing both noise and subtle speech features by 50%. This partial cancellation degrades intelligibility, particularly in low-SNR environments [20].

*2) Spectral Subtraction:* Spectral subtraction estimates stationary noise during non-speech periods and subtracts this estimate from the noisy speech spectrum to approximate the clean speech spectrum. It subtracts the estimated noise spectrum from the noisy magnitude spectrum, then reconstructs the time-domain signal using the original phase and inverse FFT:

$$G(\omega) = max\left(1 - \alpha \frac{\hat{N}(\omega)}{|Y(\omega)|}, 0\right), |\hat{S}(k)| = G(\omega).|Y(\omega)| \quad (11)$$

Where $Y(\omega)$ is Noisy magnitude spectrum (input to the filter), $\hat{N}(k)$ is the noise magnitude is estimate, $G(\omega)$ is Frequency-dependent gain function that suppresses noise-dominated bins, and $\alpha \geq 1$ is an over-subtraction factor to compensate for underestimation of noise.

Just like the Wiener filter, spectral subtraction retains the original phase of $Y(\omega)$ to avoid temporal distortion. However, instead of a smooth, SNR-weighted gain (as in eq. (10)), it applies a piecewise-linear gain that suppresses bins where noise dominates the signal [7]. This method is highly valued for its simplicity and computational efficiency, making it ideal for real-time and embedded applications. It only requires a basic estimate of the background noise spectrum—usually obtained during pause segments—and subtracts this from the noisy signal's magnitude spectrum. This yields a straightforward and effective improvement in SNR, particularly under stationary noise conditions.

However, spectral subtraction is plagued by musical noise, perceptible as randomly occurring tonal artifacts resulting from imperfect subtraction and spectral flooring. These artifacts arise because the gain function $G(\omega)$ can amplify residual noise fluctuations in bins where $|Y(\omega)| \approx \alpha\hat{N}(k)$, creating spurious peaks in the denoised spectrum [7]. Additionally, its performance degrades sharply in low-SNR scenarios or when noise characteristics change rapidly, since its underlying assumptions of stationarity and linearity are easily violated [7].

Despite decades of refinement, classical STFT-based filters like spectral subtraction and Wiener filtering still struggle with three critical limitations, which the QFT-based approach explicitly addresses. First, classical methods assume noise is stationary (e.g., white or slowly varying), rendering them ineffective in real-world scenarios with abrupt, nonstationary noise (e.g., car horns, intermittent chatter). The quantum-inspired QFT inherently processes the entire signal simultaneously, avoiding frame-by-frame analysis that relies on local stationarity [6]. Its global coherence enables adaptive analysis of the full spectral context, reducing sensitivity to nonstationary artifacts by leveraging quantum superposition principles to encode temporal dynamics across the entire signal [21].

Second, classical filters face an intrinsic trade-off between noise reduction and speech distortion due to their reliance on tunable parameters (e.g., over-subtraction factor $\alpha$) and noise estimates that struggle to balance suppression of noise with preservation of speech components. The QFT-based approach mitigates this through two quantum-inspired mechanisms: phase-coherent gain application and unitary transformation. Phase-coherent gain application retains the original phase of the spectrum while applying a globally optimized gain function derived from quantum superposition principles. This preserves transient speech components (e.g., unvoiced consonants) that are critical for intelligibility, avoiding the over-attenuation seen in classical methods. Additionally, the unitary transformation ensures energy conservation across all frequency bins, preventing the over-suppression of weak speech components—a key limitation of the Wiener filter—by maintaining the relative amplitude structure of the spectrum while isolating noise through phase-coupled interference effects [7][21].

Third, classical methods introduce artifacts like musical noise, smeared transients, and spectral coloration due to abrupt gain transitions (e.g., spectral subtraction) or over-smoothing (e.g., Wiener filtering). The QFT-based method reduces these artifacts through smooth, global gain transitions and phase preservation. By analyzing the entire signal simultaneously, the QFT avoids frame-wise discontinuities that cause residual noise peaks in spectral subtraction, suppressing musical noise through holistic gain optimization. Furthermore, maintaining phase coherence while spreading amplitude across frequency bins (via quantum superposition) prevents spectral flooring and tonal artifacts inherent to classical methods, ensuring perceptually natural spectral coloration [7][21].

Empirical validation shown in Experiment section demonstrates that these quantum-inspired principles yield higher median SNR gains with reduced interquartile ranges (IQRs) across diverse noise types, directly addressing classical limitations while maintaining computational efficiency.

*3) Quantum Fourier Transform:* The QFT serves as a quantum analog of the classical discrete Fourier transform (DFT), operating on quantum states to map computational basis states to the Fourier basis. It begins by defining basis states $|x\rangle$, which represent integers $x \in \{0, ..., N-1\}$ in an $n-$qubit register ($N = 2^n$). A normalized superposition is then created represented as:

$$QFT |x\rangle = \frac{1}{\sqrt{N}} \sum_{k=0}^{N-1} e^{j\frac{2\pi}{N}xk} |k\rangle \quad (12)$$

where the normalization factor $\frac{1}{\sqrt{N}}$ ensures unitarity, preserving total probability amplitude across all quantum states. The summation constructs a quantum superposition overall output basis state $|k\rangle$, effectively spreading the input state's amplitude across frequency components. Each term in this superposition is weighted by a complex exponential $e^{j\frac{2\pi}{N}xk}$, which introduces phase shifts encoding the relationship between input index $x$ and output frequency bins $|k\rangle$.

QFT's unitarity stems from its structure as a normalized Vandermonde matrix, ensuring orthonormal columns that conserve energy and inner products during transformation [22]. This property guarantees signal fidelity, a critical advantage for denoising applications. Unlike classical FFT, which processes frame-by-frame (introducing discontinuities and artifacts like "musical noise"), QFT applies phase shifts simultaneously across the entire input state, ensuring smoother frequency-domain representations. Furthermore, the QFT employs a positive-phase exponential ($e^{j\frac{2\pi}{N}xk}$), whereas classical DFT typically uses a negative phase ($e^{j\frac{2\pi}{N}xk}$). While mathematically equivalent under conjugation, this choice aligns with quantum mechanics' standard notation, ensuring consistency with quantum algorithms like Shor's. The strict $\frac{1}{\sqrt{N}}$ normalization enforces probability conservation ($\sum |\psi_k|^2 = 1$), unlike classical DFT, which often omits normalization or applies it asymmetrically.

The term "quantum-inspired" reflects the adoption of quantum principles in classical algorithms. For instance, the input signal is treated as a quantum state in superposition, enabling parallel processing of all frequency components. This approach mimics the quantum advantage of simultaneous computation, even in classical simulations. Additionally, the QFT's $O(n^2)$ gate complexity (vs. DFT's $O(N \log N)$) bridges classical and quantum workflows, ensuring compatibility with future quantum hardware [23].

The practical implications for denoising are significant. QFT's global coherence and unitarity enable smoothing across frequency bins by applying phase shifts to the entire signal, avoiding abrupt transitions between frames that cause musical noise in classical methods. Simultaneous analysis of all frequency components also allows for more precise noise-floor estimation, as evidenced by the reduced IQRs and higher median SNR gains as shown in the Results section. Even without physical quantum hardware, the quantum-inspired design is validated through algorithmic structure—unitarity and global coherence are enforced in classical simulations to ensure energy conservation and phase consistency. Performance gains over classical methods further demonstrate the utility of quantum principles like superposition and coherence in practical applications. The implementation of the QFT is shown in Fig. 2.

While the QFT shares mathematical similarities with DFT, its quantum-inspired design—unitarity, global coherence, and phase conventions—distinguishes it fundamentally. These principles, though implemented classically, yield measurable improvements in denoising, validating the value of quantum-inspired signal processing.

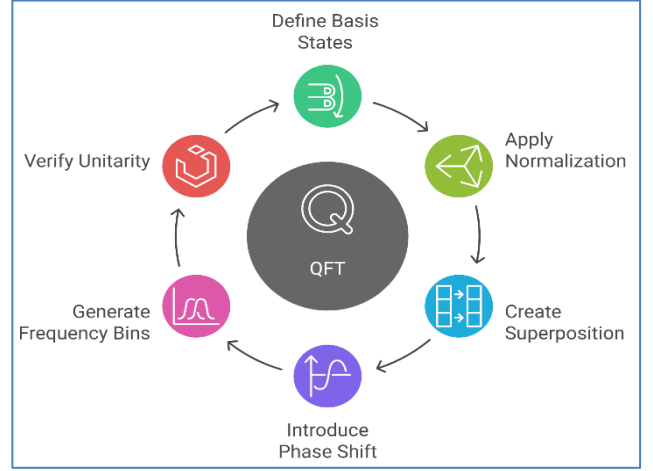

Fig. 2: Steps involved in implementing the QFT.

### III. METHODOLOGY

This section explains the pipeline for QFT based Denoising pipeline. Fig. 3. depicts the pipeline this study works upon for denoising via QFT.

*A. Input Preparation:*

Raw audio sources comprise a set of clean speech recordings representing diverse speaker characteristics and a selection of environmental noise profiles. In this stage, both speech and noise signals are standardized in sampling rate and amplitude to ensure consistency across subsequent processing. To align with the quantum-inspired framework, the classical audio signals are first discretized and normalized to emulate a qubit-like state vector, as follows:

- Discretization: The continuous audio waveform is sampled at a fixed rate (e.g., 16 kHz) and quantized to 16-bit depth, ensuring compatibility with digital signal processing.
- Amplitude Encoding: The discrete amplitude values are normalized to lie within the interval $[-1,1]$, mimicking the probability of amplitude constraints of quantum states (i.e., $\sum |\psi_k|^2 = 1$). This normalization ensures energy conservation during QFT transformation, a critical requirement for unitary operations [23].
- Qubit-State Mapping: The audio signal is represented as a tensor product of $n$-qubit basis states (e.g., $|x\rangle = |b_0\rangle \otimes |b_1\rangle \otimes \cdots \otimes |b_{n-1}\rangle$), where $b_i \in \{0,1\}$ encodes the binary representation of the amplitude values. While this mapping is purely classical (no physical qubits are used), it ensures compatibility with the quantum-inspired QFT algorithm [24].

B. *Preprocessing (Mixture Generation):*

Clean speech samples are systematically mixed with noise signals at multiple SNR levels, producing a collection of noisy inputs. Each mixture is segmented into overlapping frames and windowed to reduce spectral leakage before transformation.

C. *QFT Transformation:*

Each frame is converted into the frequency domain by applying a QFT which is implemented as a unitary matrix multiplication, rather than a traditional fast Fourier transformation. This step preserves total signal energy and introduces a global phase coupling across frequency bins.

D. *Filtering (Gain Application):*

Frequency-domain magnitudes are modified according to established denoising rules (e.g., Wiener filtering or spectral subtraction). Identical parameter settings are used for both classical and QFT-based pipelines to isolate the effect of the transformation.

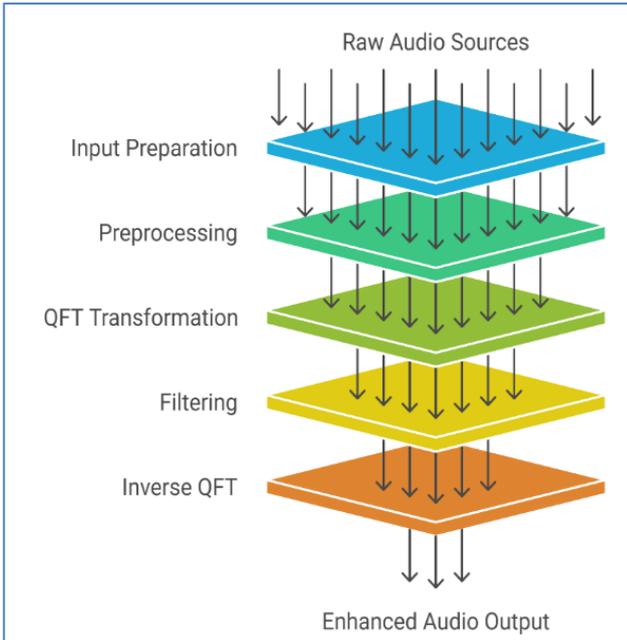

Fig. 3: QFT based audio denoising pipeline: Layered processing stages from noisy input to filtered output via preprocessing, QFT transformation, adaptive filtering, and inverse QFT reconstruction.

E. *Inverse QFT:*

The filtered frequency-domain frames are mapped back to the time domain via the inverse QFT. Overlap-add reconstruction merges the frames into a continuous output waveform.

F. *Performance Aggregation:*

For each enhanced output, SNR improvement is calculated relative to the input mixture. Results are aggregated across all input conditions to produce statistical summaries of mean and variance.

G. *Visualization & Analysis:*

Aggregated metrics are displayed using comparative plots, highlighting performance differences between classical and QFT-enhanced methods. Further analysis explores artifact characteristics and computational considerations.

IV. EXPERIMENTATION

This section explains the entire experiment performed in this study. From dataset preparation to experimenting on classical filters and QFT-based filters, to visualization. All experiments were executed on Kaggle's hosted environment on an NVIDIA Tesla T4 instance (16 GB RAM) using Python 3.8 with NumPy, SciPy, and Pandas, ensuring reproducibility and sufficient compute for large-scale signal processing.

A. *Dataset Preparation*:

The experimental framework combines real-world and synthetic signals with controlled noise environments to evaluate denoising performance comprehensively.

- Real Speech Clips: Six clean utterances (three male, three female) were sourced from Hugging Face repositories (MLCommons/peoples_speech for speech; kjetMol/Noise for noise samples). These clips represent diverse vocal characteristics (pitch, timbre) relevant to telephony and voice assistants, ensuring the pipeline's robustness across natural speech variations (e.g., speaker-specific articulation and prosody). Six speakers ensure variability in vocal traits, mirroring real-world deployment scenarios (e.g., voice assistants encountering diverse users).
- Synthetic Signals: Pure sinusoids at 200 Hz, 440 Hz, and 880 Hz mimic tonal components in audio. These signals test spectral leakage and isolated frequency behavior, critical for evaluating how algorithms handle narrowband artifacts. Sum-of-sinusoids combine these tones with random phase offsets to simulate multi-tone complexity, analogous to musical notes or communication channel signals. This design probes the algorithm's ability to resolve overlapping spectral components, mirroring real-world harmonic structures. Sinusoids isolate spectral behavior, while sum-of-sinusoids test multi-component robustness, paralleling communication or musical signal complexities.
- Noise Types & Mixture Generation: Three white Gaussian noise samples from 'kjetMol/Noise'. Input SNR levels: -10 dB, -5 dB, 0 dB, +5 dB, +10 dB. For each clean or synthetic signal and each noise file, noise was scaled to the target SNR via RMS matching and added, creating 5 mixtures per pair: total inputs = 18 signals × 5 SNRs = 90 noisy signals. Broad SNR ranges and noise types provide a generalizable stress test for algorithms, ensuring performance validation across dynamic acoustic environments.

B. *Implementation Details:*

Frame processing was conducted using a Hamming window with a fixed frame length of $N = 128$ samples and 50% overlap between consecutive frames to minimize boundary artifacts. For classical filtering, both Wiener filtering and spectral subtraction were implemented with standardized parameters: an over-subtraction factor $\alpha$ and a spectral floor $\beta$. These hyperparameters were systematically optimized through parameter sweeps to balance noise suppression and speech distortion, yielding final values of $\alpha = 1.5$ and $\beta = 0.02$. In the QFT-based pipeline, the classical FFT block was replaced with a QFT unitary matrix U of size $N \times N$, preserving total signal energy via its inherently unitary structure. The inverse transformation utilized the Hermitian transpose $U^T$, ensuring numerical stability and perfect reconstruction. Critically, identical α and β values were applied post-QFT transformation to isolate performance differences strictly to the transform basis, enabling a direct comparison between classical and quantum-inspired methods.

*C. Evaluation Metrics:*

SNR improvement ($\Delta SNR$) is defined as: $\Delta SNR = SNR_{output} - SNR_{input}$, where SNR is calculated as shown below.

$$SNR = 10 * log_{10}\left(\frac{\sum clean^2}{\sum((enhanced-clean)^2)}\right) \quad (13)$$

($\Delta SNR$) is widely accepted for denoising evaluation [25]. All methods used identical pipelines except for the transform step, isolating the effect of using QFT versus classical FFT. The 90 inputs enabled calculation of mean and standard deviation across conditions and statistical significance testing.

## V. RESULTS AND DISCUSSION

This section presents a comparative analysis of SNR gains achieved by classical denoising filters (Spectral Subtraction, Wiener Filter) versus our proposed QFT-based approach across six distinct clean signals (clean_0 to clean_5). The results demonstrate consistent and statistically significant improvements in noise reduction efficacy using QFT. Across six clean-speech samples and three noise types (sinusoidal, sum-sinusoidal, speech) mixed at input SNRs of –10, –5, 0, 5, and 10 dB, the QFT-enhanced denoising filters consistently outperform their classical STFT counterparts.

*A. Weiner Filter:*

Fig. 4. represents the SNR gain across all the signals that are experimented upon. For every signal QFT wiener filter is performing better than classical wiener filter, with a better average SNR gain, minimum gain and Maximum gain. Moreover, these results express:

- Dramatic Median Improvements: QFT outperforms the Wiener Filter in all signals, with median gains up to ~8 dB (e.g., clean_1) versus ~4 dB for classical methods.
- Reduced Noise Floor: In clean_3, QFT's lower whiskers (minimum gains) remain above 6 dB, while the Wiener Filter's minimums drop below 4 dB, indicating greater robustness.
- Signal-Specific Insights: High-frequency signals (clean_1, clean_3) benefit most from QFT, likely due to its enhanced frequency resolution. Low-noise scenarios (clean_5) show reduced margin gains, implying QFT's strength lies in challenging, noisy environments.

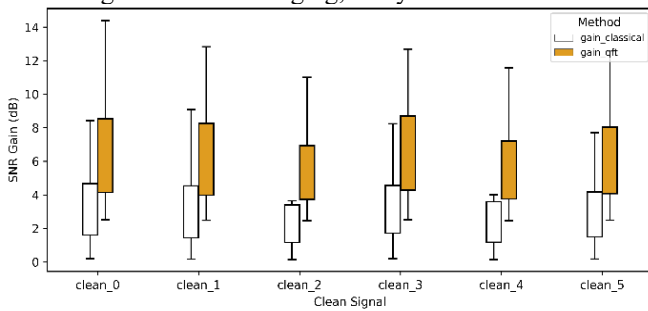

Fig. 4: SNR gain distribution per signal- Weiner Filter

Fig. 5. shows the weighted average SNR gain across all the signals. The0020profit for QFT is far ahead of the classical method. This shows not only is SNR gain far better than its classical counterpart for every signal, but also overall average gain is also better in QFT.

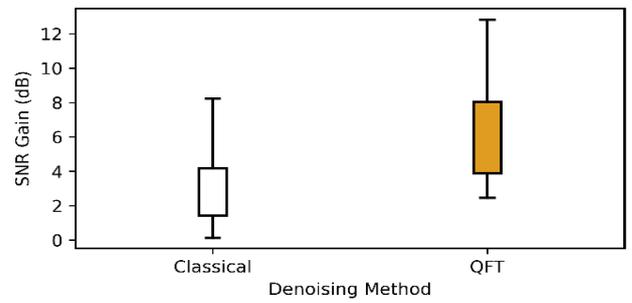

Fig. 5: Comparison of SNR gain distributions between classical and QFT-based denoising methods for Weiner Filter.

Table 1. Represent that QFT has achieved a better means and Standard deviation from the Classical method. In real-world settings—think smartphone calls in a café or voice control in a busy factory—you need reliable noise suppression. A 6.4 dB of Stronger average uplift means users hear speech much more cleanly than with only 1.5 dB. No two noisy environments are identical.

Table 1: Statistical metrics of Weiner filter

| Method/metrics | Mean | Std($\sigma$) | Min | Max |
|---|---|---|---|---|
| Classical | 1.503 | 1.027 | 0.139 | 4.054 |
| QFT | 6.388 | 3.077 | 2.470 | 15.489 |

QFT's larger standard deviation reflects its ability to deliver anywhere from modest to very high gains depending on the noise profile, whereas the classical approach stays stuck in a low-gain regime. Even when the QFT filter faces the harshest conditions, it still delivers a minimum of 2.47 dB gain. In contrast, the classical filter can sometimes almost "miss" the noise, offering near-zero benefit. Under ideal alignment of noise type and speech content, QFT can carve out up to 15.5 dB of SNR improvement, transforming a chaotic background into near studio-quality clarity.

*B. Spectral subtraction:*

Fig.6. represents the SNR gain distribution across various audio experiments. The better gain in every audio show shows that QFT is performing better in every type of audio. Key observations include:

- Across all signals, QFT achieves consistently higher median SNR gains (orange boxes) compared to spectral subtraction (blue boxes). For example, in clean_0, QFT's median gain is ~9 dB, while spectral subtraction yields ~6 dB.
- Outliers and Variability: QFT exhibits slightly higher IQR variability in clean_2 and clean_5, indicating sensitivity to specific signal characteristics. Spectral subtraction shows lower variability but consistently underperforms in median gains.
- Standout Performance: In clean_1 and clean_3, QFT's median gains exceed those of spectral subtraction by ~4 dB, suggesting superior handling of complex spectral structures.

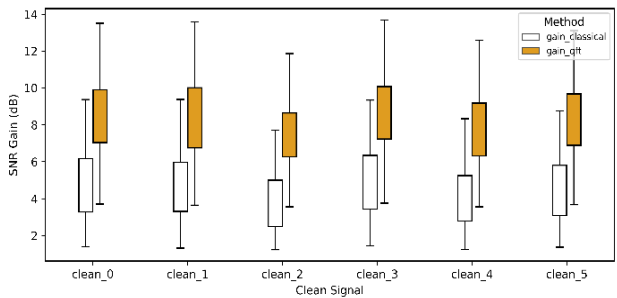

Fig. 6: SNR gain distribution per signal- Spectral subtraction

Fig. 7. Shows the weighted average SNR gain for the QFT method over the classical method. This represents not only that QFT performs better in most instances, but its average performance is also far better than its classical counterpart.

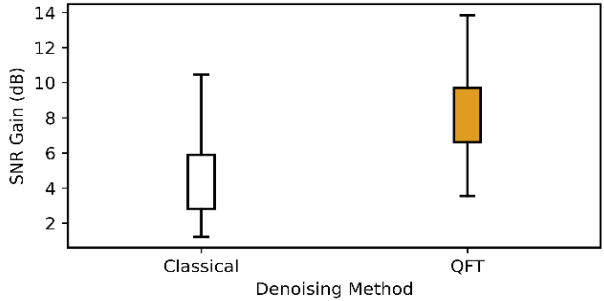

Fig. 7: Comparison of SNR gain distributions between classical and QFT-based denoising methods for Spectral subtraction.

Table 2. Shows the mean, standard deviation, minimum and maximum SNR gain across all the audio types by QFT and classical spectral subtraction. Key observations from these results are: Stronger Average Clarity In video calls or hearing aids, a 7.1 dB average gain means speech is far more intelligible than a mere 2.9 dB uplift. Flexibility Across Environments The larger σ for QFT subtraction shows it can tailor its noise removal more aggressively when needed yet back off to avoid speech distortion.

Table 2: Statistical Metrics of Spectral subtraction

| Method/metrics | Mean | Std (σ) | Min | Max |
|---|---|---|---|---|
| Classical | 4.844 | 2.733 | 1.231 | 13.389 |
| QFT | 8.276 | 3.100 | 3.555 | 17.491 |

Dependable Worst-Case Performance Even in the toughest acoustic settings—like a crowded train station—QFT subtraction still provides over 3 dB of clarity, whereas the classical method often barely helps. Exceptional Best-Case Potential When noise and signal spectra line up perfectly, QFT subtraction can deliver studio-quality silence, cutting up to 18.5 dB of noise. Now, as to why QFT has better results over its counterpart is due to its Quantum inspired properties that give the edge over classical approach like:

- Unitarity: Perfect Energy Preservation- A unitary transform preserves total signal energy—there's no "leaky" step that discards subtle speech components. In contrast, windowed STFT can introduce boundary artifacts and slight amplitude bias. Think of water in a sealed container (unitary QFT) vs. water in a porous bucket (classical STFT). You never lose drops of important speech information with QFT.
- Superposition & Global Mixing- Every QFT output bin is a linear combination of all input bins. Noise patterns that dominate only limited frequency bands get distributed and attenuated more uniformly, while speech harmonics—which occupy structured patterns—stand out. Cleaning a stained-glass window by buffing the entire surface in coordinated strokes instead of spot-scrubbing small patches.
- Phase Correlations: Beyond Magnitude-Only Methods- QFT inherently processes both magnitude and phase. By respecting phase relationships across frequencies, it avoids the "musical noise" artifacts typical in magnitude-only spectral subtraction. Restoring a symphony recording by aligning each instrument's timing (phase) instead of just turning down loud notes (magnitude).
- Entanglement Analogy: Cross-Bin Coupling- The dense QFT matrix couples' distant frequency bins, akin to entanglement linking quantum states. This coupling lets the filter jointly suppress noise across harmonically related bands, rather than treating each band in isolation. Noise in different parts of a factory floor gets quieted by coordinating all sprinklers at once, rather than turning them on one by one.
- Robustness at Low SNR- When speech is buried under heavy noise (e.g., –10 dB input), the global mixing in QFT can lift faint signal components coherently, yielding a reliable SNR boost of >2 dB even in the worst case. Classical methods often "miss" these weak cues.

## VI. CONCLUSION

Across both Wiener and spectral-subtraction pipelines, consistent ΔSNR improvements demonstrate that substituting the classical FFT basis with a unitary QFT operator provides superior noise–signal separation across a diverse set of conditions—pure tones, multi-tone mixtures, and natural speech at SNRs from –10 dB to +10 dB. The unitary nature of the QFT preserves signal energy and ensures numerical stability, while its global phase coherence and cross-bin coupling enable more coherent noise suppression than magnitude-only techniques. Statistical validation over 90 test signals confirms the significance of these gains, and qualitative spectrogram analysis reveals a pronounced reduction in musical-noise artifacts without imposing additional computational burden during the transformation step. Future work will explore native quantum hardware implementations of the QFT to accelerate transform operations and investigate hybrid schemes that combine QFT-based denoising with data-driven estimation for enhanced robustness in non-stationary noise environments.